%% file: bspsiphi_prl-v1.8.tex
\documentclass[aps,prl,showpacs,twocolumn,groupedaddress]{revtex4}  % for submission
\usepackage{graphicx}  % needed for figures
\usepackage{dcolumn}   % needed for some tables
\usepackage{bm}        % for math
\usepackage{amssymb}   % for math

\input{DecayModes.tex}

\newcommand{\jp}{\ensuremath{J/\psi}}
\newcommand{\jpsimumu}{J/\psi\rightarrow\mu^+ \mu^-}
\newcommand{\psimumu}{\psi(2S)\rightarrow\mu^+ \mu^-}

\newcommand{\bsjphi}{\ensuremath{\bs\rightarrow \jp\, \phi}}
\newcommand{\bspsiphi}{\ensuremath{\bs\rightarrow \psi(2S)\, \phi}}
\newcommand{\bujpsik}{\ensuremath{\bu\rightarrow \jp\, K^\pm}}
\newcommand{\bupsik}{\ensuremath{\bu\rightarrow \psi(2S)\, K^\pm}}

\newcommand{\gev}{\mbox{\,\rm Ge\kern -0.1emV}}
\newcommand{\mev}{\mbox{\,\rm Me\kern -0.1emV}}

\begin{document} 

\hspace{5.2in} \mbox{FERMILAB-PUB-08-134-E}

\title{Relative rates of $B$ meson decays into $\psi$(2S) and $J/\psi$ mesons}

% use the official authorlist for publication
\input list_of_authors_r2.tex  % input Dzero author list
\date{May 16, 2008}

\begin{abstract}
We report on a study of the relative rates of $B$ meson decays into $\psi$(2S) 
and $J/\psi$ mesons using $1.3$ fb$^{-1}$ of $p \bar{p}$
collisions at $\sqrt{s} = 1.96$ TeV recorded by the D0 detector
operating at the Fermilab Tevatron Collider.
We observe the channels $B^0_s\rightarrow \psi$(2S)$\phi$,
 $B^0_s\rightarrow J/\psi\phi$, $B^\pm\rightarrow \psi$(2S)$K^\pm$, and $B^\pm\rightarrow J/\psi K^\pm$
 and we measure the relative branching fractions for these channels to be
$$\frac{{\cal B}(\bpf)}{{\cal B}(\bjf)} = 0.53\pm 0.10 {\text{ (stat)}} \pm 0.07
  \text{ (syst)}\pm 0.06 { \; (\cal B)},$$  
$$\frac{{\cal B}(\bupsik)}{{\cal B}(\bujpsik)} = 0.63 \pm 0.05 {\text{ (stat)}} \pm 0.03 {\text{ (syst)}} \pm 0.07 {\; (\cal B)},$$
where the final error corresponds to the uncertainty in the $J/\psi$ and $\psi$(2S) branching ratio into two muons.
\end{abstract} 

% activate the following line for publication 

\maketitle

$B$ meson decays into final states containing charmonium play
 a crucial role in the study of CP violation and the 
precise measurement of neutral $B$ meson mixing parameters~\cite{bib:bigi-sanda}.
 For CP violation in $B_s$ mixing, $B_s\rightarrow J/\psi \phi$ decays are being
 used to measure the width difference between the mass eigenstates and the CP violating relative phase difference between the off-diagonal 
elements $\Gamma_{12}$ and $M_{12}$ describing the mixing of neutral $B$ mesons~\cite{bib:cdf-psi,bib:dzero-psi}.
%the mass difference $\Delta m_s$, and $\Delta \Gamma_s$~\cite{bib:cdf-psi,bib:dzero-psi}.  
Since the current experimental results are limited by statistics, 
it is important to establish new channels like the decay $B^0_s\rightarrow \psi$(2S)$\phi$
where these studies can be performed.  

The study of $B$ meson decays into 
several charmonium states can also be used to constrain the long-distance 
parameters associated with color octet production which are important for 
the understanding of both mixing induced and direct CP violation~\cite{bib:beneke}.
While these modes have been precisely measured in $B^+$ 
and $B^0$ decays~\cite{pdg}, 
an extension of these studies into the $B^0_s$ system
 provides an important test of quark-hadron duality.

In this Letter we report measurements of $B$ meson decays into
charmonium using the channels
$B^+\rightarrow J/\psi K^+$, $B^+\rightarrow \psi$(2S)$K^+$,
$B^0_s\rightarrow J/\psi \phi$ and $B^0_s\rightarrow \psi$(2S)$\phi$. 
Charge conjugation is implied throughout.
The $J/\psi$ and $\psi$(2S) mesons are reconstructed in the dimuon channel and the $\phi$ 
is reconstructed in the $K^+K^-$ channel. The study uses a data sample of  $p\bar{p}$ collisions at
$\sqrt{s}=1.96$ TeV corresponding to an integrated luminosity of  
approximately $1.3$ fb$^{-1}$ recorded by the D0
 detector operating at the Fermilab Tevatron Collider.
 Similar studies have recently been reported by the CDF
 collaboration~\cite{bib:cdf-psi2s}.

D0 is a general purpose detector described in detail in Ref.~\cite{run2det}.
Charged particles are reconstructed using a silicon vertex tracker
 and a scintillating fiber tracker located inside a
superconducting solenoidal coil that provides a magnetic field of approximately $2$ T.
The tracking volume is surrounded by a LAr-U calorimeter.
Muons are reconstructed using a
spectrometer consisting of magnetized iron toroids and three super-layers of
proportional tubes and plastic trigger scintillators located outside the calorimeter. 
Only data recorded by dimuon triggers were used for
this analysis. 

The selection requirements are determined using simulated samples for the four decay modes. The {\sc pythia}~\cite{bib:pythia} Monte Carlo (MC) generator is used to model $b\bar{b}$ production and fragmentation, followed by {\sc evtgen}~\cite{bib:evtgen} to simulate the kinematics of $B$-meson decay. The detector response is simulated using a {\sc geant}~\cite{bib:geant} based MC. Simulated events are processed through the same reconstruction code as used for the data.
The dimuon trigger is modeled using a detailed simulation program incorporating all aspects of the trigger logic.  The trigger simulation is verified using a data sample collected with single muon triggers. Backgrounds are modeled using data in the mass sideband regions around the candidate $B$ meson.

Muon candidates are required to have track segments reconstructed in at
least two out of the three muon system super-layers and to be associated with a track
reconstructed with hits in both the silicon and fiber trackers. 
We require that the muon transverse momentum, $p_T$, is greater than $2$ GeV$/c$.  
The charmonium system is formed
by combining two oppositely charged muon candidates that are associated
with the same track jet~\cite{bib:jet} and form a well
reconstructed vertex with $\chi^2$/DOF $< 16$. 
We require the dimuon $p_T$ to be greater than $4$ GeV$/c$.
%The resulting dimuon mass distribution is shown in Fig.~\ref{fig:mmumu}. 
The invariant mass of the dimuon system is required to be within $250$ MeV/$c^2$ of the nominal charmonium state mass~\cite{pdg}, since the invariant mass resolution is about $\approx$ 75 MeV/$c^2$.
We then redetermine the muon momenta with a mass constraint imposed when forming the $B$ meson candidate.

All charged particles within the same track jet as the dimuon system are considered as kaon candidates and the kaon mass is assigned.  
The candidates are required to have hits in both the silicon 
and fiber trackers and have $p_T> 0.6$ GeV$/c$.
Pairs of oppositely charged kaons with $p_T> 0.9$ GeV$/c$ are 
combined to form $\phi$ candidates. The expected invariant mass resolution for the $\phi$ mesons is 4~MeV/$c^2$.
Therefore the pair of kaons must form a well reconstructed vertex 
and have $1.008 < m(K^+K^-) < 1.032$ GeV$/c^2$. 

The charmonium candidates are combined with either a 
kaon or $\phi$ candidate to form either a $B^+$ or a $B^0_s$ candidate.
 The $B$ meson daughter particles are required to 
form a well reconstructed vertex and have an 
invariant mass between $4.4$ and $6.2$ GeV$/c^2$. 
% We require the vertex $\chi^2$ to be less than 20 
%for 3 DOF for $B^+$ candidates and less than 36 
%for 5 DOF for $B^0_s$ candidates.

Backgrounds from prompt charmonium production are reduced by 
requiring the $B$ meson decay vertex to be displaced from the 
interaction point in the transverse plane by more than 
4 times the error on the measured displacement for $B^+$
 candidates and 6 times the error for $B^0_s$ candidates. 
For all candidates, the error on the displacement measurement 
is required to be less than $150$~$\mu$m.  
Combinatorial backgrounds are reduced by requiring the 
$B$ candidate momentum vector to be aligned with the position vector 
of the secondary vertex to within 26 degrees. Possible background contamination from kaons and pions misidentified as muons and other source of B decays which could result in a peaking background have been studied and found to be negligible.

\begin{figure}[h]
\begin{center}
  \includegraphics*[width=\linewidth]{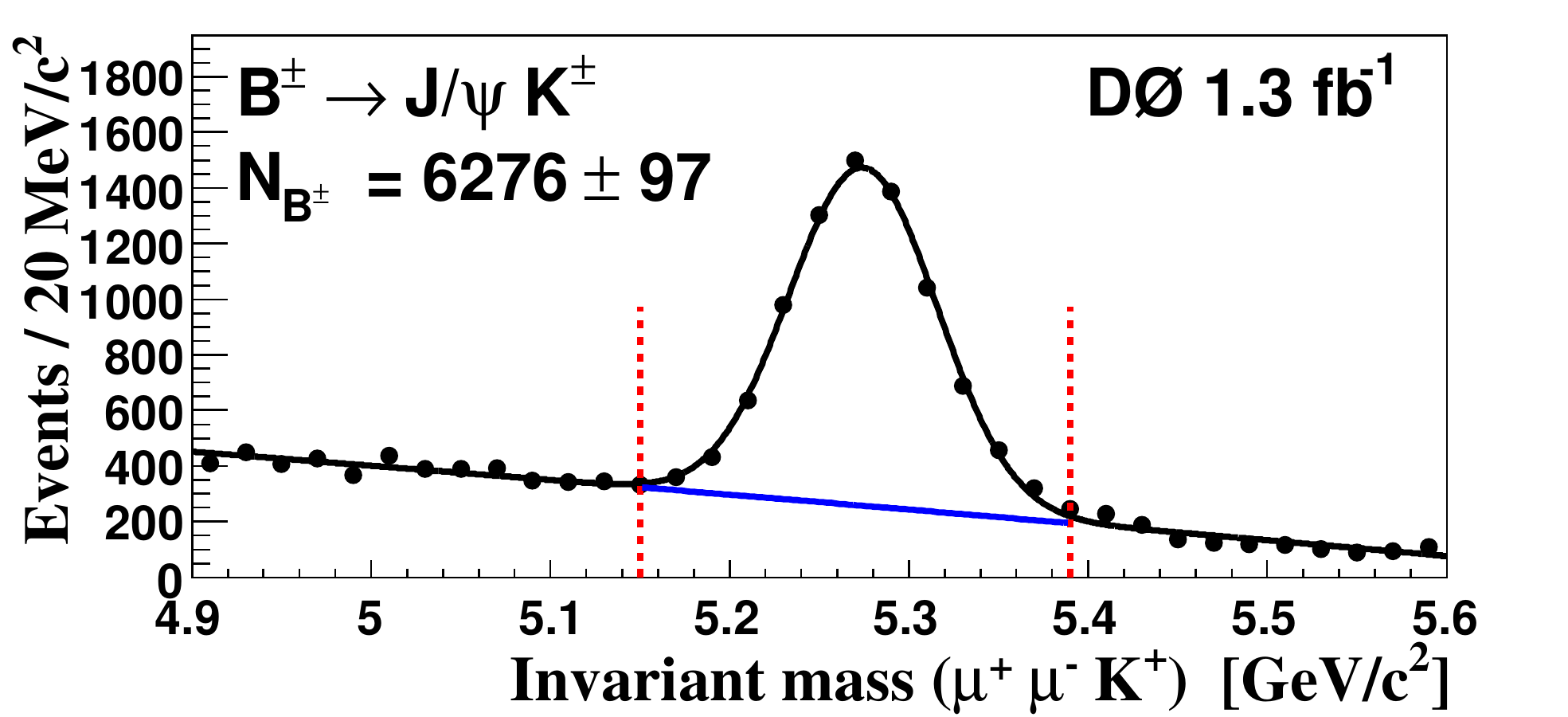}
  \caption{\label{fig1} 
  $\mu^+ \mu^- K^\pm$ invariant mass distribution for the
  $\bujpsik$ data selection. The region between two dashed vertical lines represents the signal window.}
\end{center}
\end{figure}

\begin{figure}[h]
\begin{center}
  \includegraphics*[width=\linewidth]{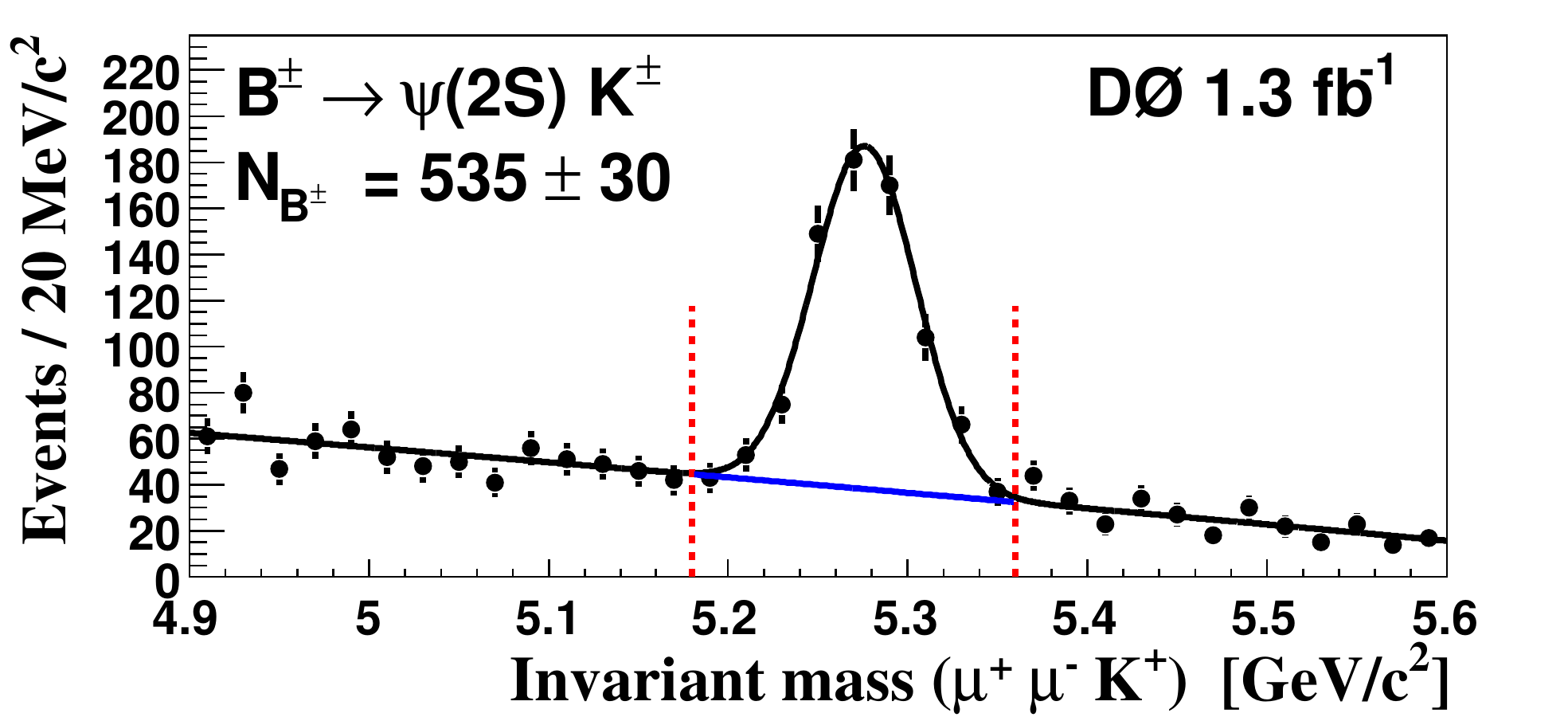}
  \caption{\label{fig2}
  $\mu^+ \mu^- K^\pm$ invariant mass distribution for the
  $\bupsik$ data selection. The region between two dashed vertical lines represents the signal window.}
\end{center}
\end{figure}

\begin{figure}[h]
\begin{center}
  \includegraphics*[width=\linewidth]{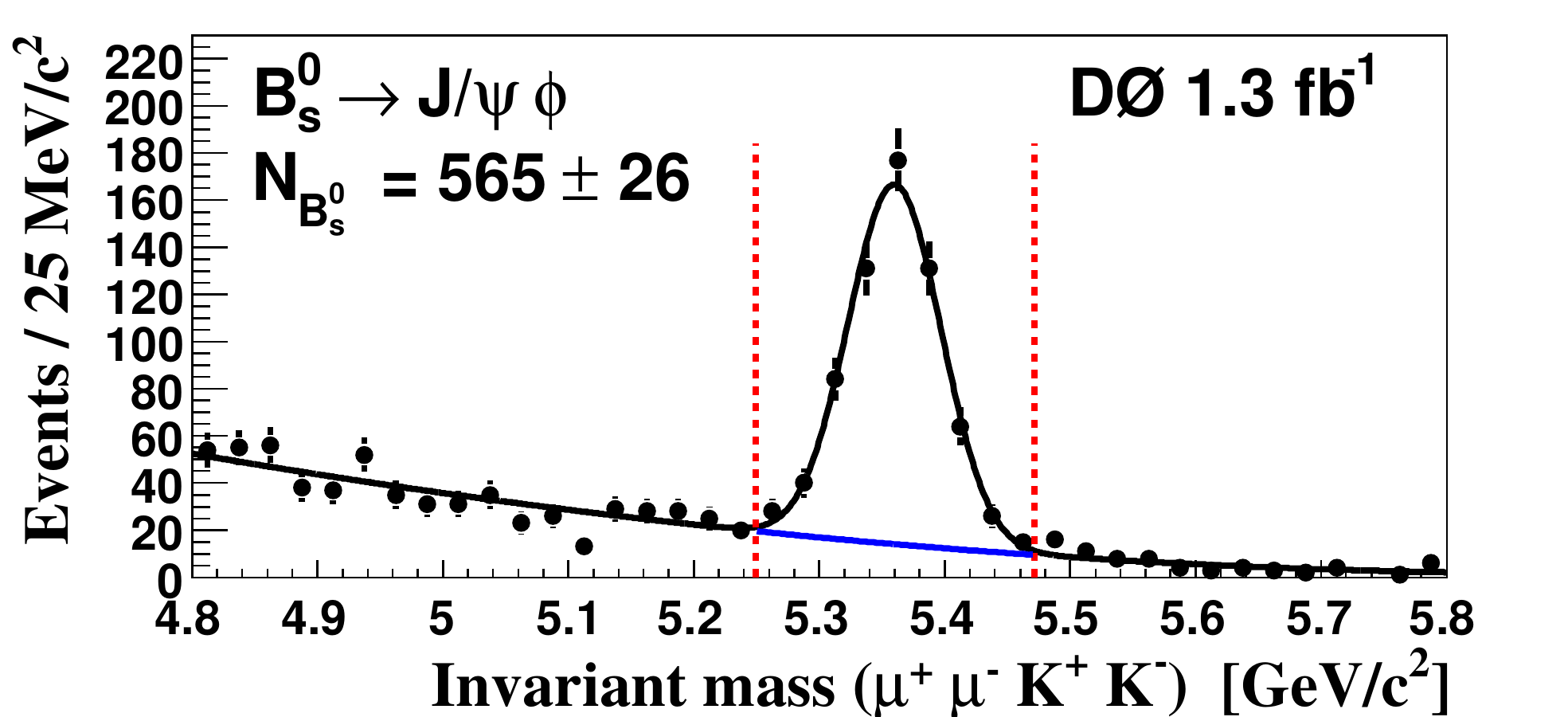}
  \caption{\label{fig4} 
  $\mu^+ \mu^- \phi$
  invariant mass distribution for the
  $B^0_s \rightarrow J/\psi\, \phi$ data selection. The region between two dashed vertical lines represents the signal window.}
\end{center}
\end{figure}

\begin{figure}[h]
\begin{center}
  \includegraphics*[width=\linewidth]{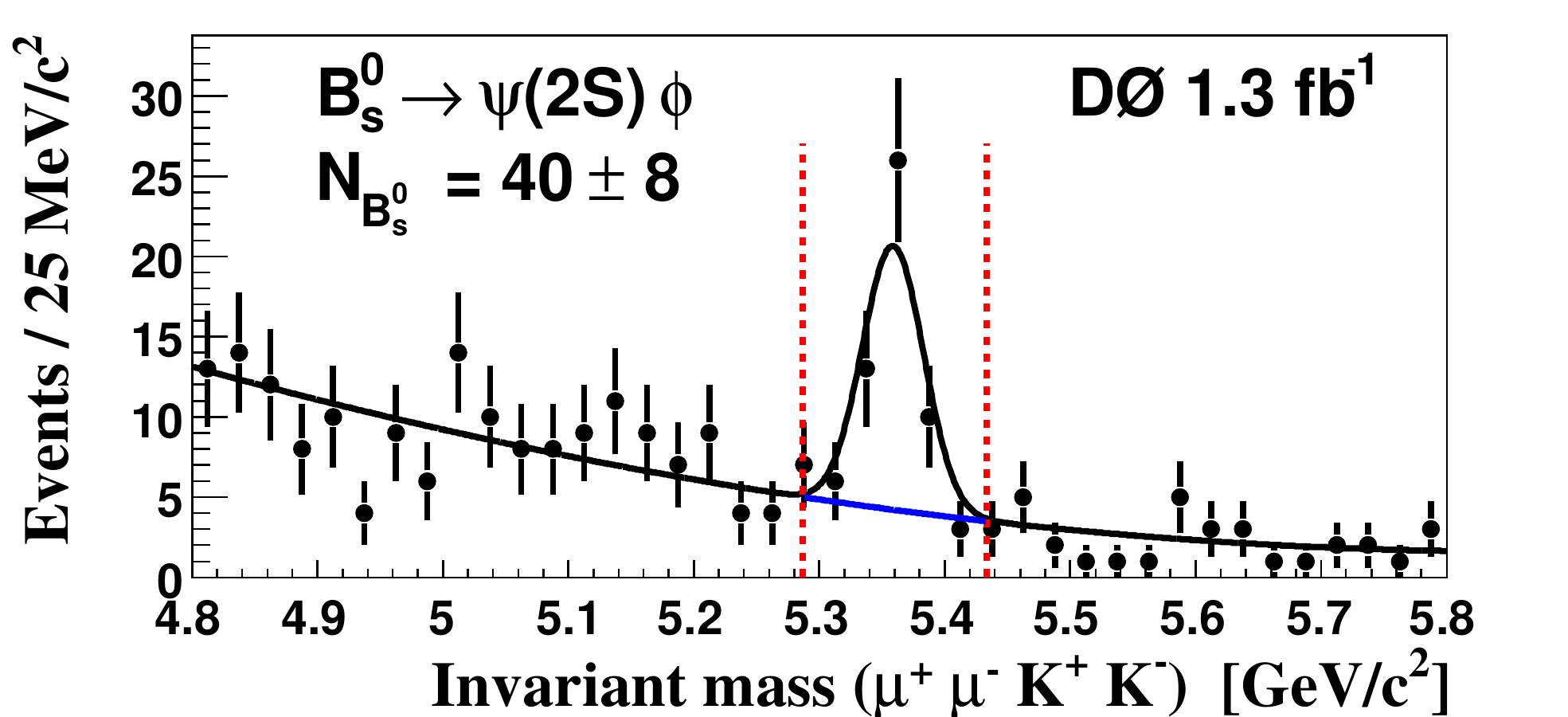}
  \caption{\label{fig3}
  $\mu^+ \mu^- \phi$
  invariant mass distribution for the
   $\bspsiphi$ data selection. The region between two dashed vertical lines represents the signal window.}
\end{center}
\end{figure}

The resulting mass distributions of the $B$ meson candidates are displayed 
in Figs.~\ref{fig1}-\ref{fig3}. The $B$ meson yield 
is extracted from the data using a binned likelihood fit %$\chi^2$ fit 
to the data assuming a Gaussian component for signal
and a second-order polynominal distribution for background.  
The number of signal events is obtained by integrating the fit functions over the range of interest. This range is indicated by the two dashed vertical lines in Figs.~\ref{fig1}-\ref{fig3} for each channel, and covers a 
a region corresponding $\pm 3\sigma$, where $\sigma$ is the expected invariant mass resolution. We see signals in all four channels. The results of the signal yields, corrected for the background contributions, are listed in Table~\ref{tab:fit_results}.


\begin{center} 
\begin{table}[h] 
\small 
\begin{center} 
  \caption{\label{tab:fit_results}Summary of obtained event yields from the fits as described in the text and signal efficiencies obtained from MC simulations.}
  \begin{tabular}{lcc}\hline\hline 
   Decay & Efficiency $$ & Yield\\ \hline
$B^\pm \rightarrow J/\psi K^+$ &  $(1.07 \pm 0.02) \cdot 10^{-3}$  & 6276$\pm$97 \\
$B^\pm \rightarrow \psi(2S) K^+$  &  $(1.14 \pm 0.04) \cdot 10^{-3}$ &535$\pm$30 \\ \hline
$B^0_s \rightarrow J/\psi\, \phi$ &  $(14.4 \pm 0.7) \cdot 10^{-5}$  & 565$\pm$26 \\
$B^0_s \rightarrow \psi(2S)\, \phi$   &   $(15.2 \pm 0.6) \cdot 10^{-5}$ & 40$\pm$8\\\hline\hline
 \end{tabular} 
\end{center} 
\end{table} 
\end{center} 

The relative yield of $B$ meson decays into $\psi$(2S) and $J/\psi$ mesons is given by 
\begin{eqnarray}\label{brratio1}
\lefteqn{\frac{{\cal B}(B\rightarrow \psi{\rm (2S)}M)}{{\cal B}(B\rightarrow J/\psi M)} \,= }\\ \nonumber
& & \frac{N_{\psi{\rm (2S)}M}}{N_{J/\psi M}}\cdot 
  \frac{\epsilon_{J/\psi M}}{\epsilon_{\psi{\rm (2S)} M}} \cdot \frac{{\cal B}(\jpsimumu)}{{\cal B}(\psimumu)},
\end{eqnarray} 
where $B$ is either a $B^+$ or $B^0_s$ meson, $M$ is either a $K^+$ or $\phi$ meson,
 $N$ is the number of signal events returned from the fit, and $\epsilon$ 
is the reconstruction efficiency determined from MC.  
The measured branching fractions ${\cal B}( J/\psi 
  \rightarrow \mu^+ \mu^-)\, = \, (5.93 \pm 0.06) \cdot 10^{-2}$ and ${\cal B}( \psi(2S) \rightarrow \mu^+ 
  \mu^-)\, =\, (7.5\pm 0.8) \cdot 10^{-3}$ are taken from Ref.~\cite{pdg} and combined into a ratio of branching fractions  ${\cal B}( J/\psi \rightarrow \mu^+ \mu^-)\, /\, {\cal B}( \psi(2S) \rightarrow \mu^+ \mu^-)\, =\,8.12 \pm 0.89$.
  The uncertainty on the ratio is given by the uncertainty on the single measured 
  branching fractions assuming no correlations. 
%All numbers are listed in Table~\ref{tab:numbers}.

For the measurement of $B$ meson branching fractions the sources of 
systematic uncertainties are (i)
%\begin{itemize} 
%\item 
the branching fractions of the charmonium mesons to dimuons, 
%\item
(ii) systematics of the individual signal yield determinations and
%\item 
(iii) the determination of the efficiencies 
  $\epsilon_{\psi(2S)\phi}$ and $\epsilon_{J/\psi \phi}$. In the ratio 
%  one expects most effects to cancel out. This is due to the fact that 
 % both decay modes have very similar topologies. 
 many systematic uncertainties cancel, such as the integrated luminosity, $b$ production and fragmentation and the selection efficiencies. 
  However, the polarization could be different for the $B^0_s$ 
  decays. We use a pure CP-even state for the generated 
$B^0_s\rightarrow \psi(2S)\phi$ and $B^0_s\rightarrow J/\psi\phi$ MC.
%   afor the normalization channel however also a pure CP odd state was generated. 
%\end{itemize} 

The relative uncertainties that enter into the calculation of the 
relative branching fractions are given in Table~\ref{sys_b}. 
The uncertainty related to the measured charmonium resonance branching fractions 
enter both measurements and are the same for both. The uncertainties are treated as uncorrelated 
and give a combined uncertainty of 11\% on each of the ratios of branching fractions.

The relative statistical uncertainties on the efficiencies 
$\epsilon_{\psi(2S)\, K}$ and $\epsilon_{J/\psi\, K}$ are 3.7\% 
and 1.9\%, respectively. They are combined into a single statistical uncertainty on the efficiency ratio 
assuming no correlations. To obtain and estimate the signal yield variation,
the background shape and fit range for the background region are varied. This
yields a variation of 3\% for $B^\pm\rightarrow \psi(2S)\,K^\pm$ and
0.9\% for $B^\pm\rightarrow J/\psi \ \,K^\pm$.
The ratio of branching fractions ${\cal B}(B^\pm\rightarrow \psi(2S)K^\pm)/{\cal B}(B^\pm\rightarrow J/\psi K^\pm)$ is then  0.63 $\pm$ 0.05 (stat) $\pm$ 0.03 (syst) $\pm$ 0.07 ($\cal B$).

The relative uncertainties on 
$\epsilon_{\psi(2S)\phi}$ and $\epsilon_{J/\psi\phi}$ are 4.5\% 
and 5.6\%, respectively. These uncertainties are combined into a single statistical uncertainty on the efficiency ratio assuming no correlations. As an estimate of the signal yield variation, the 
shape of the background as well as the invariant mass regions for the background estimations are changed. This gives a variation of 7.5\% for the $\bpf$ and 2.1\% for the $\bjf$ signal yield.
The $B^0_s\rightarrow J/\psi\phi$ and $B^0_s\rightarrow \psi(2S) \phi$ MC events are generated as pure CP-even decays with a $\bs$ lifetime of
1.44~ps~\cite{Abazov:2004ce}. To account for a possible efficiency
difference related with the different lifetime of the $\bs$, the
$B^0_s\rightarrow J/\psi\phi$ MC events are weighted according to the combined world average
lifetime~\cite{Group(HFAG):2005rb}. The efficiency difference
is estimated to be 8\%, which is taken as an additional systematic
uncertainty. The resulting ratio of branching fractions is ${\cal B}(\bpf)/{\cal
    B}(\bjf))$ = 0.53 $\pm$ 0.10 (stat) $\pm$ 0.07 (syst) $\pm$ 0.06 ($\cal B$).
\begin{center} 
\begin{table}[h] 
\small 
\begin{center} 
\caption{\label{sys_b}Relative systematic uncertainties on the 
  ratio of branching fractions ${\cal B}(B^\pm\rightarrow \psi(2S)\,K^\pm)/{\cal 
    B}(B^\pm\rightarrow J/\psi\,K^\pm)$ and ${\cal B}(B^0_s\rightarrow \psi(2S)\phi)$/${\cal B}(B^0_s\rightarrow J/\psi \phi)$.
}
  \begin{tabular}{lcc}\hline\hline 
    Source & \multicolumn{2}{c}{Relative Uncertainty $[\%]$} \\ \hline
	& $(J/\psi, \psi(2S)) K^\pm$&  $(J/\psi, \psi(2S)) \phi$\\ \hline 
    ${\cal B}( J/\psi \rightarrow \mu \mu )$ & \multicolumn{2}{c}{1.7} \\ 
    ${\cal B}( \psi(2S) \rightarrow \mu \mu )$ & \multicolumn{2}{c}{11} \\ \hline
    Total ${\cal B}$ &  \multicolumn{2}{c}{11} \\\hline 
    $\epsilon_{J/\psi (K^\pm,\phi)}/\epsilon_{\psi(2s) (K^\pm,\phi)}$  & \multicolumn{1}{c}{4.1}&  7.2\\ 
     Event yield  $\psi(2S)$ channel             & 3.0  & 7.5 \\ 
     Event yield $J/\psi$ channel & 0.9 & 2.1\\
    CP odd-even mixture ($J/\psi \phi$)      & N.A.          & 8.0 \\  \hline
    Total (syst) & 5.2  & 13.3 \\\hline \hline 
\end{tabular} 
\end{center} 
\end{table} 
\end{center} 

In summary, we have presented the observation of the decay $B^0_s\rightarrow \psi(2S)\phi$
with the decay $\psi(2S) \ra \mu^+ \mu^-$ at D0 and performed a measurement of the ratio of branching fractions
\begin{eqnarray}\label{resultbs}
\lefteqn{\frac{{\cal B}(\bspsiphi)}{{\cal B}(\bsjphi)} \,= }\\ \nonumber
& & 0.53\pm 0.10 {\text{ (stat)}} \pm 0.07 {\text{ (syst)}} \pm 0.06 \; ( \cal B).
\end{eqnarray} 
In addition, a measurement of the ratio of branching fractions 
\begin{eqnarray}\label{resultbu}
\lefteqn{\frac{{\cal B}(\bupsik)}{{\cal B}(\bujpsik)} \,= }\\ \nonumber
& & 0.63 \pm 0.05 {\text { (stat)}} \pm 0.03 {\text{ (syst)}} \pm 0.07 \;(\cal B)
\end{eqnarray} 	
has been performed. These results are competitive and in good agreement with published measurements~
\cite{Aubert:2001xs,bib:cdf-psi2s}. The combination with these measurements should result in a significant precision improvement on the measured ratios of branching fractions.

\input acknowledgement_paragraph_r2.tex
\end{document}

%% file: DecayModes.tex
\newcommand{\ra}{\ensuremath{\rightarrow}}
\newcommand{\bs}{\ensuremath{B_s^0}}             % Bs
\newcommand{\bu}{\ensuremath{B^{\pm}}}               % B+
\newcommand{\bjf}{\ensuremath{\bs\ra J/\psi \phi}}
\newcommand{\bpf}{\ensuremath{\bs\ra \psi(2S) \phi}}

%% file: list_of_authors_r2.tex
% LIST_OF_AUTHORS_R2.TEX               3/27/08              
%
\author{V.M.~Abazov$^{36}$}
\author{B.~Abbott$^{75}$}
\author{M.~Abolins$^{65}$}
\author{B.S.~Acharya$^{29}$}
\author{M.~Adams$^{51}$}
\author{T.~Adams$^{49}$}
\author{E.~Aguilo$^{6}$}
\author{S.H.~Ahn$^{31}$}
\author{M.~Ahsan$^{59}$}
\author{G.D.~Alexeev$^{36}$}
% ...

\affiliation{\vspace{0.1 in}(The D\O\ Collaboration)\vspace{0.1 in}}
% ...
\affiliation{$^{6}$University of Alberta, Edmonton, Alberta, Canada,
                Simon Fraser University, Burnaby, British Columbia, Canada,
                York University, Toronto, Ontario, Canada, and
                McGill University, Montreal, Quebec, Canada}
% ...
\affiliation{$^{29}$Tata Institute of Fundamental Research, Mumbai, India}
% ...
\affiliation{$^{31}$Korea Detector Laboratory, Korea University, Seoul, Korea}
% ...
\affiliation{$^{36}$Joint Institute for Nuclear Research, Dubna, Russia}
% ...
\affiliation{$^{49}$Florida State University, Tallahassee, Florida 32306, USA}
% ...
\affiliation{$^{51}$University of Illinois at Chicago,
                Chicago, Illinois 60607, USA}
% ...
\affiliation{$^{59}$Kansas State University, Manhattan, Kansas 66506, USA}
% ...
\affiliation{$^{65}$Michigan State University,
                East Lansing, Michigan 48824, USA}
% ...
\affiliation{$^{75}$University of Oklahoma, Norman, Oklahoma 73019, USA}
% ...

%% file: acknowledgement_paragraph_r2.tex
% acknowledgement_paragraph_r2.tex                         3/27/08
%
We thank the staffs at Fermilab and collaborating institutions, 
and acknowledge support from the 
DOE and NSF (USA);
CEA and CNRS/IN2P3 (France);
FASI, Rosatom and RFBR (Russia);
CNPq, FAPERJ, FAPESP and FUNDUNESP (Brazil);
DAE and DST (India);
Colciencias (Colombia);
CONACyT (Mexico);
KRF and KOSEF (Korea);
CONICET and UBACyT (Argentina);
FOM (The Netherlands);
STFC (United Kingdom);
MSMT and GACR (Czech Republic);
CRC Program, CFI, NSERC and WestGrid Project (Canada);
BMBF and DFG (Germany);
SFI (Ireland);
The Swedish Research Council (Sweden);
CAS and CNSF (China);
and the
Alexander von Humboldt Foundation.